\documentclass[conference]{IEEEtran}
\usepackage{fancyhdr}

\usepackage{epsfig,latexsym}
\usepackage{float}
\usepackage{indentfirst}
\usepackage{amsmath}
\usepackage{amssymb}
\usepackage{times}
\usepackage{subfigure}
\usepackage{pifont}
\usepackage{psfrag}
\usepackage{cite}
\usepackage{url}
\usepackage{lastpage}
\usepackage{bm}
\usepackage{color}
\usepackage{fancyhdr}
\usepackage[center]{caption2}
\usepackage{balance}

\newtheorem{Proposition}{Proposition}

\ifCLASSINFOpdf  
\else
\fi

\begin{document}
%
\title{ Secrecy Rate Maximization for MISO Multicasting \\SWIPT System with Power Splitting Scheme}

\author{\IEEEauthorblockN{Miao Zhang，, Kanapathippillai Cumanan and Alister Burr }
\IEEEauthorblockA{Department of Electronics, University of York, York, United Kingdom\\
Email: \{mz1022, kanapathippillai.cumanan, alister.burr\}@york.ac.uk}}


%


\maketitle


\begin{abstract}
 This paper considers transmit covariance matrix design for secrecy rate maximization problem in a multiple-input single-output (MISO) multicasting simultaneous wireless information and power transfer (SWIPT) system. In order to enhance the performance of the system, artificial noise (AN) is added to the transmit signal in the design for the following purposes: to reduce the received signal-to-noise ratio (SNR) at the eavesdroppers and increase the harvested energy. We assume that all the channel-state-information (CSI) is perfectly known at the transmitter and all legitimate users are capable of simultaneously receiving information and harvesting energy. In addition, all the eavesdroppers are passive and they can harvest energy only when they are not intercepting or eavesdropping the messages intended for the legitimate users. The original secrecy rate maximization problem is not convex in terms of transmit and artificial covariance matrices as well as the power splitting (PS) ratio. In order to circumvent this non-convexity issue, we exploit the \emph{Charnes-Cooper} Transformation and semidefinite relaxation (SDR) to convert this original problem into a convex one. However, this convex problem does not always yield the rank-one transmit and AN covariance matrices to obtain the solution of the original problem. Therefore, we analyze the optimal conditions and utilize a Gaussian randomization (GR) method to construct the rank-one solutions from the non-rank one results. Simulation results have been provided to demonstrate the performance of the proposed transmit covariance matrices design for MISO multicasting SWIPT system.
\end{abstract}

%
\IEEEpeerreviewmaketitle

\section{Introduction}
Information security is one of the most challenging problem in wireless communications due to the open access of wireless medium. The security schemes in traditional wireless networks are developed in upper layers through cryptographic encryptions \cite{liang,cumanan2012new,cumanan2009sinr}. The broadcast nature of wireless communications introduces different challenges in terms of key exchange and distributions \cite{chu1,cumanan2014secrecy,chu2017robust,cumanan2010multiuser,cumanan2010sinr}. In order to improve the security performance of wireless transmissions, the information-theoretic based physical layer security was first proposed by Shannon \cite{Shannon}, and then developed by Wyner \cite{Wyner} and Csiszar \cite{csiszar}. This approach implements security in the physical layer to complement the cryptographic methods by exploiting the channel state information (CSI) of legitimate parties and eavesdroppers. In \cite{Wyner}, Wyner introduced the wiretap channel model and the concept of secrecy capacity, which is defined as the difference of the mutual information between the legitimate channel and the wiretap channel.\\
\indent Recently, simultaneous wireless information and power transfer (SWIPT) has significantly influenced the research community as the radio signals can be exploited to harvest energy for power constrained devices including unreachable sensor nodes. Traditional energy harvesting schemes utilize the energy from nature such as hydro-power, solar and wind. However, these methods are not suitable for mobile terminals, since the aspects of geographical position, climate and the design of devices may limit the performance of energy harvesting \cite{ng2015secure}. In order to circumvent these issues, wireless power transfer has been recently proposed as the solution and has become an interesting research area as evidenced by the recent work in the literature \cite{ng2015secure,varshney,grover2010shannon}. In \cite{liu}, secrecy wireless information and power transfer based beamforming design has been proposed for a SWIPT system to avoid the eavesdropping by the energy receivers through physical layer security. On the other hand, the information receivers could also exploit wireless energy harvesting technique to simultaneously receive the information and energy through a power splitting (PS) scheme \cite{yuan2014joint}, where the received signal is divided into information and energy streams \cite{leng2014power}. However, this could introduce some security threats to information receivers as the received signal strength at energy receivers is better than that of the information receivers and the achieved secrecy rate decreases to zero \cite{liu}. To address this issue, we consider secrecy rate maximization for a multiple-input single-output (MISO) multicasting SWIPT with PS scheme with minimum energy requirements at legitimate users and energy receivers. In particular, transmit beamforming vectors are designed to maximize the secrecy rate and satisfy the energy constraints based on convex optimization approach \cite{boyd2004convex}. \\
\indent In this paper, we consider a MISO secure multicasting SWIPT system, where the transmitter and legitimate users with multiple and single antennas, respectively. Similar to \cite{chu2015robust}, the eavesdroppers are equipped with mutiple antennas. In addition, it is assumed that the transmitter has the perfect channel state information (CSI) of all links. The energy receivers could attempt to intercept the information intended for the legitimate users and could turn out to be the potential eavesdroppers in the network. In particular, both the legitimate users and the eavesdroppers are able to decode information and harvest energy simultaneously. However, in order to guarantee confidential transmission, we focus on the worst scenario that the PS ratio for information decoding at the eavesdroppers are assumed to one and the transmitter only guarantees the harvested energy requirements at the eavesdroppers when they do not attempt to eavesdrop \cite{leng2014power}. To ensure the secure communication between legitimate terminals and satisfy the energy requirements, we exploit artificial noise (AN) approach to confuse the eavesdroppers or degrade the decoding capability while providing the required energy \cite{chu1}\cite{chu2016secrecy}. For this network set up, we formulate the beamforming design into a secrecy rate maximization problem with energy constraints. This original problem is not convex in terms of beamforming vectors. To circumvent this non-convexity issue, we exploit semidefinite relaxation (SDR) \cite{luo2010semidefinite} and \emph{Charnes-Cooper} Transformation \cite{charnes1962programming} techniques to cast the problem into a semidefinite programming (SDP), which can be efficiently solved through interior point methods \cite{grant2009cvx}.\\
\indent The remainder of this paper is organized as follows. The system model is presented in Section II, while the secrecy rate maximization problem is formulated in section III. Section IV provides simulation results to validate the performance of the proposed design and Section V concludes this paper. 
\subsection{Notations}
We use the upper case boldface letters for matrices and lower boldface for vectors. $I$ denotes the identity matrix whereas $(\cdot)^{-1}$, $(\cdot)^T$ and $(\cdot)^H$ stand for inverse, transpose and conjugate transpose operation, respectively. $\mathbf{A}\succeq\mathbf{0}$ means that $\mathbf{A}$ is a positive semidefinite matrix. $|\mathbf{A}|$ and $||\mathbf{A}||$ represents the determinant and the Euclidean norm of matrix $\mathbf{A}$, respectively. The $\textrm{rank}(\mathbf{A})$ denotes the rank of a matrix, and $\textrm{tr}(\mathbf{A})$ represents the trace of matrix $\mathbf{A}$. The circularly symmetric complex Gaussian (CSCG) distribution is represented by $\mathcal{CN}(\mu,\sigma^2)$ with mean $\mu$ and variance $\sigma^2$. $\mathbb{H}^{N}$ denotes the set of all $N \times N$ Hermitian matrices.

\section{System Model}
We consider a secure multicasting MISO SWIPT system, where the legitimate transmitter establishes secured communication links with $K$ legitimate users in the presence of $L$ multiple antenna eavesdroppers. Here, it is assumed that both legitimate users and eavesdroppers are employed with the PS scheme to simultaneously decode the information and harvest the energy. In addition, the transmitter is equipped with $N_{T}$ transmit antennas and each legitimate user consists of single antenna whereas all eavesdroppers are equipped with $N_{E}$ receive antennas. The channel coefficients between the transmitter and the $k$-th legitimate user as well as the $l$-th eavesdropper are denoted by $\mathbf{h}_{s,k}\in\mathcal{C}^{N_{T}\times1}$ and $\mathbf{H}_{e,l}\in\mathcal{C}^{N_{T}\times N_{E}}$, respectively. Thus, the received signal at the $k$-th legitimate user and the $l$-th eavesdropper can be expressed as
\begin{equation}
y_{s,k}=\mathbf{h}_{s,k}^{H}\mathbf{x}+n_{sa,k},~k=1,2,...,K
\end{equation}
\begin{equation}
y_{e,l}=\mathbf{H}_{e,l}^{H}\mathbf{x}+\mathbf{n}_{ea,l},~l=1,2,...,L
\end{equation}
where $\mathbf{x}\in\mathcal{C}^{N_{T}\times1}$ denotes the transmitted signal, which can be written as
 $\mathbf{x}=\mathbf{q}s+\mathbf{v}$, 
 where $\mathbf{q}\in\mathcal{C}^{N_{T}\times1}$ is the transmit beamforming, $s$ is the information signal and $\mathbf{v}\in\mathcal{C}^{N_{T}\times1}$ is the AN. In addition, all of the receivers exploit PS to handle the received signal, then we can write (1) and (2) as
\begin{equation}
y_{s,k}=\sqrt{\rho_{s,k}}(\mathbf{h}_{sa,k}^{H}\mathbf{x}+n_{sa,k})+n_{sp,k},~\forall k
\end{equation}
\begin{equation}
y_{e,l}=\sqrt{\rho_{e,l}}(\mathbf{H}_{ea,l}^{H}\mathbf{x}+\mathbf{n_{ea,l}})+\mathbf{n}_{ep,l}, ~\forall l
\end{equation}
where $\rho_{s,k}\in(0,1]$ and $\rho_{e,l}\in(0,1]$ denote the PS factor of the $k$-th legitimate user and the $l$-th eavesdropper, respectively. The antenna noise at the $k$-th legitimate receiver and the $l$-th eavesdropper are represented by $n_{sa,k}\sim\mathcal{CN}(0,\sigma_{sa,k}^{2})$ and $n_{ea,l}\in\mathcal{C}^{N_{E}\times1}\sim\mathcal{CN}(0,\sigma_{ea,l}^2\mathbf{I})$, respectively, whereas $n_{sp,k}\sim\mathcal{CN}(0,\sigma_{sp,k}^2)$ and $\mathbf{n}_{ep,l}\in\mathcal{C}^{N_{E}\times1}\sim\mathcal{CN}(0,\sigma_{ep,l}^{2}\mathbf{I})$ represent the signal processing noise for the $k$-th legitimate receiver and the $l$-th eavesdropper, respectively. In this paper, we model the AN vector as covariance matrix, where $\mathbf{V}=\mathbf{v}\mathbf{v}^{H}$, $\mathbf{V}\in\mathbb{H}^{N_{T}}$, $\mathbf{V}\succeq 0$.

The mutual information of the $k$-th legitimate user can be written as
\begin{equation}
R_{s,k}=\log_{2}\bigg(1+\frac{\mathbf{h}_{s,k}^{H}\mathbf{q}\mathbf{q}^{H}\mathbf{h}_{s,k}}{\mathbf{h}_{s,k}^{H}\mathbf{V}\mathbf{h}_{s,k}+\sigma_{sa,k}^{2}+\frac{\sigma_{sp,k}^{2}}{\rho_{s,k}}}\bigg)
\end{equation}
and the mutual information of the $l$-th eavesdropper is written as
\begin{align}
&R_{e,l}=\nonumber\\
&\log_{2}\!\bigg|\mathbf{I}\!+\![\rho_{e,l}(\sigma_{ea,l}^{2}\mathbf{I}\!+\!\mathbf{H}_{e,l}^{H}\mathbf{V}\mathbf{H}_{e,l})\!+\!\sigma_{ep,l}^{2}\mathbf{I}]^{\!-\!1}\!\rho_{e,l}\mathbf{H}_{e,l}^{H}\mathbf{q}\mathbf{q}^{H}\mathbf{H}_{e,l}\bigg|\nonumber\\
&\leq\log_{2}\bigg|\mathbf{I}\!+\!(\mathbf{H}_{e,l}^{H}\mathbf{q}\mathbf{q}^{H}\mathbf{H}_{e,l})^{-1}\rho_{e,l}\mathbf{H}_{e,l}^{H}\mathbf{q}\mathbf{q}^{H}\mathbf{H}_{e,l}\bigg|\!=\!R_{e,l}^{\textrm{UP}}
\end{align}
The upper bound is obtained by setting $\rho_{e,l}=1$ and $\sigma_{e,l}^{2}=\sigma_{ea,l}^{2}+\sigma_{ep,l}^{2}$. Here, $\eta_{s,k}\in(0,1]$ and $\eta_{e,l}\in(0,1]$ are the power transformation ratio of the $k$-th legitimate user and the $l$-th energy receiver, respectively. The harvested power at the $k$-th legitimate user can be written as 
\begin{equation}
E_{s,k}=\eta_{s,k}(1-\rho_{s,k})\bigg[\mathbf{h}_{s,k}^{H}\mathbf{q}\mathbf{q}^{H}\mathbf{h}_{s,k}+\mathbf{h}_{s,k}^{H}\mathbf{V}\mathbf{h}_{s,k}+\sigma_{sa,k}^{2}\bigg]
\end{equation}
whereas the harvested power at the $l$-th energy receiver can be written as
\begin{align}
E_{e,l}\!\!=\!\!\eta_{e,l}(1\!\!-\!\!\rho_{e,l})\!\bigg[\textrm{tr}(\mathbf{H}_{e,l}^{H}\mathbf{q}\mathbf{q}^{H}\mathbf{H}_{e,l})\!\!+\!\!\textrm{tr}(\mathbf{H}_{e,l}^{H}\mathbf{V}\mathbf{H}_{e,l}\!\!+\!\!N_{E}\sigma_{ea,l}^{2})\bigg]
\end{align}
\section{Problem Formulation}
Here, we consider the secrecy rate maximization problem for this multicasting MISO SWIPT network, where the minimum secrecy rate between the legitimate users is maximized with transmit power and energy harvesting constraints. This problem can be formulated as 
\begin{subequations}\label{eq:Sec_rate_max_ori}
\begin{align}
\max_{\mathbf{q},\mathbf{V},\rho_{s,k}} & \min_{k,l} R_{k} = R_{s,k} - R_{e,l}^{\textrm{UP}}  \label{eq:Sec_rate_max_obj} \\
s.t. &~ \min_{k} E_{s,k} \geq \bar{E}_{s}, ~\min_{l} E_{e,l} \geq \bar{E}_{e},~ \forall k,l, \label{eq:Sec_rate_max_EH_contraints}\\
&~ \|\mathbf{q}\|^{2} + \textrm{tr}(\mathbf{V}) \leq P_{\textrm{total}}, \label{eq:Sec_rate_max_power_constraints}\\
&~ 0 < \rho_{s,k} \leq 1, \mathbf{V} \succeq \mathbf{0}. \label{eq:Sec_rate_max_another_constraints}
\end{align}
\end{subequations}

The physical meaning of the constraint in \eqref{eq:Sec_rate_max_EH_contraints} is that the transmitter should satisfy the minimum power requirement at the $l$-th passive eavesdropper if it is only interested in energy harvesting and not in eavesdropping (i.e., $\rho_{e,l}=0$). For convenience, the power transformation ratio is assumed to be $\eta_{s,k}=\eta_{e,l}=1$ and this problem can be expressed by introducing the transmit covariance matrix $\mathbf{Q}_{s}=\mathbf{q}\mathbf{q}^{H}$ as
\begin{subequations}
	\begin{align}
	\max_{\mathbf{Q}_{s},\mathbf{V},\rho_{s,k},t}& \min_{k}  \log_{2}\bigg(1\!+\!\frac{\mathbf{h}_{s,k}^{H}\mathbf{Q}_{s}\mathbf{h}_{s,k}}{\mathbf{h}_{s,k}^{H}\mathbf{V}\mathbf{h}_{s,k}\!+\!\sigma_{sa,k}^{2}\! +\!\frac{\sigma_{sp,k}^{2}}{\rho_{s,k}}}\bigg)\!+\! \log_{2}( t ) \label{eq:Sec_rate_max_obj_slack_variable_t}\\
	\!\!\!\!s.t.~ & \log_{2}\! \bigg| \mathbf{I}\!+\! (\mathbf{H}_{e,l}^{H}\mathbf{V}\mathbf{H}_{e,l}\!+\! \sigma_{e,l}^{2}\mathbf{I})\!^{\!-\!1} \mathbf{H}_{e,l}^{H}\mathbf{Q}_{s}\mathbf{H}_{e,l} \bigg|\!\leq\!\log(\frac{1}{t}), \label{eq:Eves_rate_constraint}\\
	&(1\!-\!\rho_{s,k})\bigg[ \mathbf{h}_{s,k}^{H}\mathbf{Q}_{s}\mathbf{h}_{s,k}  \!+\! \mathbf{h}_{s,k}^{H}\mathbf{V}\mathbf{h}_{s,k} \!+\!\sigma_{sa,k}^{2} \bigg]\! \geq\! \bar{E}_{s}, \label{eq:User_EH_constraint} \\
	 &\textrm{tr}(\mathbf{H}_{e,l}^{H}\mathbf{Q}_{s}\mathbf{H}_{e,l})\! +\! \textrm{tr}(\mathbf{H}_{e,l}^{H}\mathbf{V}\mathbf{H}_{e,l}) \!+\! N_{E}\sigma_{ea,l}^{2}\! \geq\! \bar{E}_{e}, \label{eq:Eve_EH_constraint} \\
	&\textrm{tr}(\mathbf{Q}_{s})+\textrm{tr}(\mathbf{V}) \leq P_{\textrm{total}},\\
	& 0 \!<\! \rho_{s,k} \!\leq \!1, \mathbf{Q}_{s}\!\succeq \!\mathbf{0},\mathbf{V}\! \succeq\! \mathbf{0},\textrm{rank}(\mathbf{Q}_{s})\! =\! 1.
	\end{align}
\end{subequations}
The constraint in \eqref{eq:Eves_rate_constraint} can be recast by removing the logarithm from both sides as 
\begin{align}\label{eq:Eve_rate_LMI}
\eqref{eq:Eves_rate_constraint} \Rightarrow (t^{-1}-1) (\mathbf{H}_{e,l}^{H}\mathbf{V}\mathbf{H}_{e,l} + \sigma_{e,l}^{2}\mathbf{I}) \succeq \mathbf{H}_{e,l}^{H}\mathbf{Q}_{s}\mathbf{H}_{e,l} 
\end{align}
It can be easily seen that \eqref{eq:Eve_rate_LMI} is a linear matrix inequality (LMI) constraint \cite{boyd2004convex} whereas \eqref{eq:Eves_rate_constraint} and \eqref{eq:Eve_rate_LMI} are equivalent. Then we obtain
\begin{align}
\eqref{eq:User_EH_constraint} \Rightarrow \mathbf{h}_{s,k}^{H}\mathbf{Q}_{s}\mathbf{h}_{s,k}  + \mathbf{h}_{s,k}^{H}\mathbf{V}\mathbf{h}_{s,k} \geq \frac{\bar{E}_{s}}{1-\rho_{s,k}} - \sigma_{sa,k}^{2}
\end{align}
\begin{align}
\eqref{eq:Eve_EH_constraint} \Rightarrow \textrm{tr}(\mathbf{H}_{e,l}^{H}\mathbf{Q}_{s}\mathbf{H}_{e,l}) + \textrm{tr}(\mathbf{H}_{e,l}^{H}\mathbf{V}\mathbf{H}_{e,l}) \geq \bar{E}_{e} - N_{E}\sigma_{ea,l}^{2}
\end{align}
Therefore, the problem can be formulated as
\begin{subequations}
	\begin{align}
	\max_{\mathbf{Q}_{s},\mathbf{V},\rho_{s,k},t} & \min_{k}  \log_{2}\!\bigg(1\!+\frac{\mathbf{h}_{s,k}^{H}\mathbf{Q}_{s}\mathbf{h}_{s,k}}{\mathbf{h}_{s,k}^{H}\mathbf{V}\mathbf{h}_{s,k}\!+\!\sigma_{sa,k}^{2} \!+\!\frac{\sigma_{sp,k}^{2}}{\rho_{s,k}}}\bigg)\!+\! \log_{2}( t )  \\ 
	s.t.~ & (t^{-1} - 1) (\mathbf{H}_{e,l}^{H}\mathbf{V}\mathbf{H}_{e,l} + \sigma_{e,l}^{2}\mathbf{I}) \succeq \mathbf{H}_{e,l}^{H}\mathbf{Q}_{s}\mathbf{H}_{e,l}, \\
	& \mathbf{h}_{s,k}^{H}\mathbf{Q}_{s}\mathbf{h}_{s,k}  + \mathbf{h}_{s,k}^{H}\mathbf{V}\mathbf{h}_{s,k} \geq \frac{\bar{E}_{s}}{1-\rho_{s,k}} - \sigma_{sa,k}^{2}, \\
	& \textrm{tr}(\mathbf{H}_{e,l}^{H}\mathbf{Q}_{s}\mathbf{H}_{e,l})\! +\! \textrm{tr}(\mathbf{H}_{e,l}^{H}\mathbf{V}\mathbf{H}_{e,l}) \geq \bar{E}_{e}\!-\! N_{E}\sigma_{ea,l}^{2}, \\
	&\textrm{tr}(\mathbf{Q}_{s})+\textrm{tr}(\mathbf{V}) \leq P_{\textrm{total}},\\
	&0 \!<\! \rho_{s,k} \!\leq \!1, \mathbf{Q}_{s}\!\succeq \!\mathbf{0},\mathbf{V}\! \succeq\! \mathbf{0},\textrm{rank}(\mathbf{Q}_{s})\! =\! 1.
	\end{align}
\end{subequations}
The above problem is still not convex in terms of transmit covariance matrices as well as the PS ratio and therefore cannot be solved using existing software. To circumvent this issue, we convert the original problem into a two-level optimization problem. The outer problem can be written with respect to (w.r.t.) the variable $ t $ as 
\begin{align}\label{eq:Outer_problem}
R^{*} \!=\! \max_{t} \log_{2}(1\! +\! f(t) ) \!+\! \log_{2}(t), ~s.t. ~t_{\min} \!\leq\! t\! \leq\! 1,
\end{align}
whereas the inner problem can be expressed as 
\begin{align}\label{eq:Inner_problem}
f(t) &= \max_{\mathbf{Q}_{s},\mathbf{V},\rho_{s,k}} \min_{k} \frac{\mathbf{h}_{s,k}^{H}\mathbf{Q}_{s}\mathbf{h}_{s,k}}{\mathbf{h}_{s,k}^{H}\mathbf{V}\mathbf{h}_{s,k}+\sigma_{sa,k}^{2} +\frac{\sigma_{sp,k}^{2}}{\rho_{s,k}}} \nonumber\\
s.t. & \max_{l} \log_{2}\! \bigg| \mathbf{I}\!+\! (\mathbf{H}_{e,l}^{H}\mathbf{V}\mathbf{H}_{e,l}\!\!+\! \sigma_{e,l}^{2}\mathbf{I})\!^{-\!1}\! \mathbf{H}_{e,l}^{H}\mathbf{Q}_{s}\mathbf{H}_{e,l} \bigg|\! \leq\! \log_{2}\!(\frac{1}{t}), \nonumber\\
&~ \mathbf{h}_{s,k}^{H}\mathbf{Q}_{s}\mathbf{h}_{s,k}  + \mathbf{h}_{s,k}^{H}\mathbf{V}\mathbf{h}_{s,k} \geq \frac{\bar{E}_{s}}{1-\rho_{s,k}} - \sigma_{sa,k}^{2}, \nonumber\\
&~ \textrm{tr}(\mathbf{H}_{e,l}^{H}\mathbf{Q}_{s}\mathbf{H}_{e,l}) + \textrm{tr}(\mathbf{H}_{e,l}^{H}\mathbf{V}\mathbf{H}_{e,l}) \geq \bar{E}_{e} - N_{E}\sigma_{ea,l}^{2}, \nonumber\\
&~\textrm{tr}(\mathbf{Q}_{s})+\textrm{tr}(\mathbf{V}) \leq P_{\textrm{total}}, \nonumber\\
&~ 0 <\rho_{s,k} \leq 1, \mathbf{Q}_{s}\succeq \mathbf{0},~ \mathbf{V} \succeq \mathbf{0},~\textrm{rank}(\mathbf{Q}_{s}) = 1.
\end{align}
The upper bound of $ t $ in \eqref{eq:Outer_problem} is 1 due to \eqref{eq:Eves_rate_constraint}, and the lower bound $ t_{\min} $ can be derived from \eqref{eq:Sec_rate_max_obj_slack_variable_t} as 
\begin{align}
t & \!\geq \!\bigg(\! 1\!+\!\frac{\mathbf{h}_{s,k}^{H}\mathbf{Q}_{s}\mathbf{h}_{s,k}}{\mathbf{h}_{s,k}^{H}\mathbf{Q}_{s}\mathbf{h}_{s,k}\!+\!\sigma_{sa,k}^{2}\!+\!\frac{\sigma_{sp,k}^{2}}{\rho_{s,k}}} \bigg)^{-1}\!\!\!\!\! \geq\! \bigg(\! 1\!+\!\frac{\mathbf{h}_{s,k}^{H}\mathbf{Q}_{s}\mathbf{h}_{s,k}}{\sigma_{sa,k}^{2}} \bigg)^{-1} \nonumber\\
& \geq \bigg( 1\!+\!\frac{\lambda_{\max}(\mathbf{Q}_{s})\|\mathbf{h}_{s,k}\|^{2}}{ \sigma_{sa,k}^{2} } \bigg)^{-1} \!\geq\! \bigg( 1\!+\!\frac{\textrm{tr}(\mathbf{Q}_{s})\|\mathbf{h}_{s,k}\|^{2}}{ \sigma_{sa,k}^{2} } \bigg)^{-1} \nonumber\\
&\geq \bigg( 1+\frac{P_{\textrm{total}}\|\mathbf{h}_{s,k}\|^{2}}{ \sigma_{sa,k}^{2} } \bigg)^{-1} = t_{\min},
\end{align} 
where the last inequality is obtained from the total power constraint. The outer problem in \eqref{eq:Outer_problem} is a single-variable optimization problem with a bounded interval constraint $[t_{\min}, 1]$, which can be solved through an one-dimensional line search, provided that $ f(t) $ can be evaluated for any feasible $ t $. Therefore, in the following, we will focus on the inner problem in \eqref{eq:Inner_problem}, which is a fractional programming problem. Generally, bisection search is employed to tackle this problem. However, the complexity of this method based on one dimensional search algorithm is high and difficult to implement. The inner problem in \eqref{eq:Inner_problem} can be written as 
\begin{align}
\tilde{f}(t) &= \max_{\mathbf{Q}_{s},\mathbf{V},\rho_{s,k}} \min_{k} \frac{\mathbf{h}_{s,k}^{H}\mathbf{Q}_{s}\mathbf{h}_{s,k}}{\mathbf{h}_{s,k}^{H}\mathbf{V}\mathbf{h}_{s,k}+\sigma_{sa,k}^{2} + \frac{\sigma_{sp,k}^{2}}{\rho_{s,k}} } \nonumber\\
s.t. &~ (t^{-1}-1) (\mathbf{H}_{e,l}^{H}\mathbf{V}\mathbf{H}_{e,l} + \sigma_{e,l}^{2}\mathbf{I}) \succeq \mathbf{H}_{e,l}^{H}\mathbf{Q}_{s}\mathbf{H}_{e,l}, \nonumber\\
&~ \mathbf{h}_{s,k}^{H}\mathbf{Q}_{s}\mathbf{h}_{s,k}  + \mathbf{h}_{s,k}^{H}\mathbf{V}\mathbf{h}_{s,k} \geq \frac{\bar{E}_{s}}{1-\rho_{s,k}} - \sigma_{sa,k}^{2}, \nonumber\\
&~ \textrm{tr}(\mathbf{H}_{e,l}^{H}\mathbf{Q}_{s}\mathbf{H}_{e,l}) + \textrm{tr}(\mathbf{H}_{e,l}^{H}\mathbf{V}\mathbf{H}_{e,l}) \geq \bar{E}_{e} - N_{E}\sigma_{ea,l}^{2}, \nonumber\\
&~\textrm{tr}(\mathbf{Q}_{s})+\textrm{tr}(\mathbf{V}) \leq P_{\textrm{total}}, \nonumber\\
&~ 0 < \rho_{s,k} \leq 1, ~\mathbf{Q}_{s}\succeq \mathbf{0},~ \mathbf{V} \succeq \mathbf{0},~\textrm{rank}(\mathbf{Q}_{s}) = 1.
\end{align}
Then, we exploit \emph{Charnes-Cooper} transformation \cite{charnes1962programming}
\begin{align}
\mathbf{Q}_{s} = \frac{\mathbf{\tilde{Q}}_{s}}{\xi},~\mathbf{V} = \frac{\mathbf{\tilde{V}}}{\xi}, ~
\rho_{s,k} =\frac{\tilde{\rho}_{s,k}}{\xi},
\end{align}
and we can obtain
	\begin{align}
	\tilde{f}(t) & = \max_{\mathbf{Q}_{s},\mathbf{V},\tilde{\rho}_{s,k},\xi} \min_{k} \mathbf{h}_{s,k}^{H}\mathbf{\tilde{Q}}_{s}\mathbf{h}_{s,k}  \nonumber\\
	s.t. &~ \mathbf{h}_{s,k}^{H}\mathbf{\tilde{V}}\mathbf{h}_{s,k}+ \xi \sigma_{sa,k}^{2} + \frac{\sigma_{sp,k}^{2}}{\tilde{\rho}_{s,k}} = 1, \nonumber\\
	&~ (t^{-1}-1) (\mathbf{H}_{e,l}^{H}\mathbf{\tilde{V}}\mathbf{H}_{e,l} + \xi \sigma_{e,l}^{2}\mathbf{I}) \succeq \mathbf{H}_{e,l}^{H}\mathbf{\tilde{Q}}_{s}\mathbf{H}_{e,l}, \nonumber\\
	&~ \mathbf{h}_{s,k}^{H}\mathbf{\tilde{Q}}_{s}\mathbf{h}_{s,k}  + \mathbf{h}_{s,k}^{H}\mathbf{\tilde{V}}\mathbf{h}_{s,k} \geq \frac{\xi^{2} \bar{E}_{s}}{\xi-\tilde{\rho}_{s,k}} - \xi \sigma_{sa,k}^{2}, \nonumber\\
	&~ \textrm{tr}(\mathbf{H}_{e,l}^{H}\mathbf{\tilde{Q}}_{s}\mathbf{H}_{e,l}) + \textrm{tr}(\mathbf{H}_{e,l}^{H}\mathbf{\tilde{V}}\mathbf{H}_{e,l}) \geq \xi(\bar{E}_{e} - N_{E}\sigma_{ea,l}^{2}), \nonumber\\
	&~\textrm{tr}(\mathbf{\tilde{Q}}_{s})+\textrm{tr}(\mathbf{\tilde{V}}) \leq \xi P_{\textrm{total}}, \nonumber\\
	&~  0 < \tilde{\rho}_{s,k} \leq \xi,  \label{eq:Power_splitting_ratio_constraint}\nonumber\\ 
	&~\mathbf{\tilde{Q}}_{s}\succeq \mathbf{0},~ \mathbf{\tilde{V}} \succeq \mathbf{0},~\textrm{rank}(\mathbf{\tilde{Q}}_{s}) = 1.
	\end{align}
Note that (20) is equivalent to (18), the proof can be found in \cite{li2011}. Thus, the inner problem can be relaxed by removing the rank constranit as 
\begin{align}
\tilde{f}(t) & = \max_{\mathbf{\tilde{Q}}_{s},\mathbf{\tilde{V}},\tilde{\rho}_{s,k},\xi} \theta \nonumber\\
s.t. &~ \mathbf{h}_{s,k}^{H}\mathbf{\tilde{Q}}_{s}\mathbf{h}_{s,k} \geq \theta,\nonumber\\  
&~ \mathbf{h}_{s,k}^{H}\mathbf{\tilde{V}}\mathbf{h}_{s,k}+ \xi \sigma_{sa,k}^{2} + \frac{\sigma_{sp,k}^{2}}{\tilde{\rho}_{s,k}} = 1, \nonumber\\
&~ (t^{-1}-1) (\mathbf{H}_{e,l}^{H}\mathbf{\tilde{V}}\mathbf{H}_{e,l} + \xi \sigma_{e,l}^{2}\mathbf{I}) \succeq \mathbf{H}_{e,l}^{H}\mathbf{\tilde{Q}}_{s}\mathbf{H}_{e,l}, \nonumber\\
&~ \mathbf{h}_{s,k}^{H}\mathbf{\tilde{Q}}_{s}\mathbf{h}_{s,k}  + \mathbf{h}_{s,k}^{H}\mathbf{\tilde{V}}\mathbf{h}_{s,k} \geq  \frac{\xi^{2} \bar{E}_{s}}{\xi-\tilde{\rho}_{s,k}} - \xi \sigma_{sa,k}^{2}, \nonumber\\
&~ \textrm{tr}(\mathbf{H}_{e,l}^{H}\mathbf{\tilde{Q}}_{s}\mathbf{H}_{e,l}) + \textrm{tr}(\mathbf{H}_{e,l}^{H}\mathbf{\tilde{V}}\mathbf{H}_{e,l}) \geq \xi(\bar{E}_{e} - N_{E}\sigma_{ea,l}^{2}), \nonumber\\
&~\textrm{tr}(\mathbf{\tilde{Q}}_{s})+\textrm{tr}(\mathbf{\tilde{V}}) \leq \xi P_{\textrm{total}}, \nonumber\\
&~  0 < \tilde{\rho}_{s,k} \leq \xi, ~\mathbf{\tilde{Q}}_{s}\succeq \mathbf{0},~ \mathbf{\tilde{V}} \succeq \mathbf{0}.
\end{align}
The above problem is convex for a given $ t $ by relaxing the non-convex rank-one constraint, and can be solved by using interior-point method. \\
\begin{Proposition}\label{proposition:rank_proof}
Suppose we obtain $\mathbf{Q}^{*}_{s} = \frac{\mathbf{\tilde{Q}}^{*}_{s}}{\xi^{*}}, \mathbf{V}^{*} = \frac{\mathbf{\tilde{V}}^{*}}{\xi^{*}}$, where $\tilde{\mathbf{Q}}^{*}_{s}$, $\mathbf{\tilde{V}}^{*}$ and $\xi^{*}$ are the optimal solutions of (21). The rank of $\mathbf{Q}^{*}_{s}$ is less than or equal to $K$ (i.e., $\textrm{rank}(\mathbf{Q}^{*}_{s})\leq K$) and satisfies $\textrm{rank}^{2}(\mathbf{Q}^{*}_{s})+\textrm{rank}^{2}(\mathbf{V}^{*})\leq 2K+L$
\end{Proposition}
\begin{IEEEproof}
	 Please refer to Appendix.
\end{IEEEproof}
\vspace{0.9em}
By exploiting \emph{Proposition} 1, it is easy to show that the optimal solution to (21) returns rank-one. Thus a particular optimal solution is employed by considering rank-reduction algorithm \cite{huang2010rank}.
 Therefore, we considered two approaches to obtain the achievable secrecy rate: 1) 'SDR' approach, recover $\mathbf{Q}^{*}_{s} = \frac{\mathbf{\tilde{Q}}^{*}_{s}}{\xi^{*}}, ~\mathbf{V}^{*} = \frac{\mathbf{\tilde{V}}^{*}}{\xi^{*}}, ~
 \rho^{*}_{s,k} =\frac{\tilde{\rho}^{*}_{s,k}}{\xi{*}}$, and the achievable secrecy rate can be obtained by
 \begin{align}
 R_{ach}=&\log_{2}\bigg(1+\frac{\mathbf{h}_{s,k}^{H}\mathbf{Q}^{*}_{s}\mathbf{h}_{s,k}}{\mathbf{h}_{s,k}^{H}\mathbf{V}^{*}\mathbf{h}_{s,k}+\sigma_{sa,k}^{2}+\frac{\sigma_{sp,k}^{2}}{\rho^{*}_{s,k}}}\bigg)\nonumber\\
 &-\log_{2}\bigg|\mathbf{I}\!+\!(\mathbf{H}_{e,l}^{H}\mathbf{V}^{*}\mathbf{H}_{e,l})^{-1}\mathbf{H}_{e,l}^{H}\mathbf{Q}^{*}_{s}\mathbf{H}_{e,l}\bigg|
 \end{align}
 2) 'SDR+GR' approach, there the first step is same as the 'SDR' approach and then apply the GR technique, the details of which can be found in \cite{sidiropoulos2006transmit}. In addition, $\log_2[\tilde{f}^{*}(t)]$
 is the upper bound of the secrecy capacity to satisfy the constraints.
 
\section{Simulation Results}
\begin{center}
\begin{figure}[ht!]
\includegraphics[width=\linewidth]{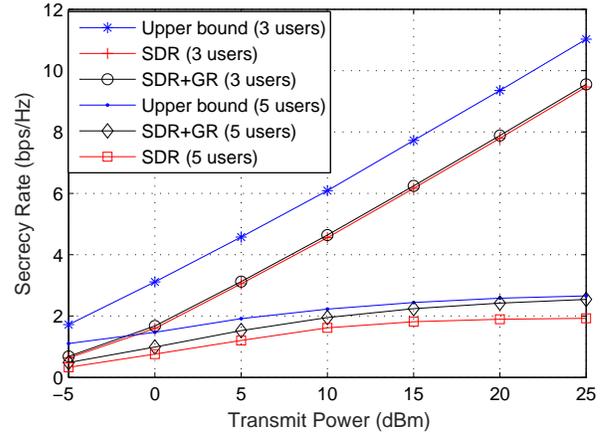}
\caption{Secrecy Rate versus Transmit Power}	
\label{fig:SRP}
\end{figure}
\end{center}

\begin{center}
	\begin{figure}[ht!]
		\includegraphics[width=\linewidth]{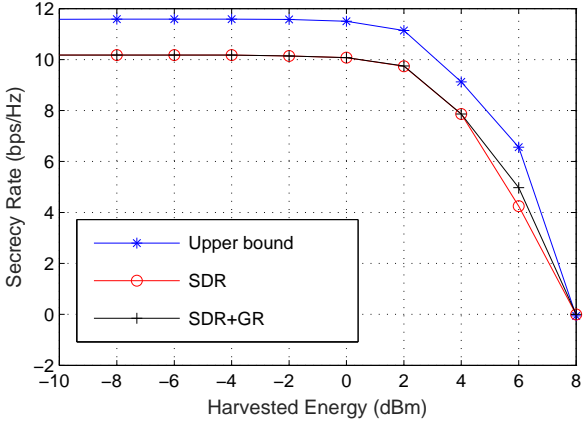}
		\caption{Secrecy Rate versus Harvested Energy}	
		\label{fig:SRE}
	\end{figure}
\end{center}

In this section, we provide numerical simulation results to validate the performance of the proposed schemes. In particular, we consider a MISO multicasting SWIPT network with different number of legitimate users (3 and 5) and three eavesdroppers.It is assumed that the transmit and all the eavesdroppers consist of five $N_{T}=5$ and two $N_{E}=2$ antennas. All the channels coefficients are generated by CSCG with zero mean and $10^{-3}$ variance. All noise variances are assumed to be $10^{-7}$ and the minimum harvested energy for all legitimate users and eavesdroppers are assumed to be equal. The legend 'Upper bound' in Fig. \ref{fig:SRP} and Fig. \ref{fig:SRE} presents the values of $\log_{2}[\tilde{f}^{*}(t)]$ where $\tilde{f}^{*}(t)$ is the optimal solution of (21). The results denoted by 'SDR' is obtained by determining $\mathbf{Q}^{*}_{s} = \frac{\mathbf{\tilde{Q}}^{*}_{s}}{\xi^{*}}, \mathbf{V}^{*} = \frac{\mathbf{\tilde{V}}^{*}}{\xi{*}},
\rho^{*}_{s,k} =\frac{\tilde{\rho}^{*}_{s,k}}{\xi{*}}$ whereas the results denoted by 'SDR+GR' are obtained by using SDR approach and GR techniques.

Fig. \ref{fig:SRP} represents the achieved secrecy rates with different transmit power and different numbers of legitimate users based on 'SDR' and 'SDR+GR' approaches. As seen in Fig. \ref{fig:SRP} the performance of 'SDR+GR' is better than that of 'SDR'. In addition, the performance gap between 'SDR+GR' and the upper bound is not significant in particular for five users scenario. 

Fig. \ref{fig:SRE} depicts the relationship between secrecy rate and the harvested energy, where the available transmit power is assumed to be 30 dBm with 3 legitimate users. As seen in Figure \ref{fig:SRE}, the harvested energy increases as the secrecy rate decreases. The achievable secrecy rate significantly reduces when the energy harvesting requirement is larger than 2 dBm and it is zero at 8 dBm energy harvesting. However, the performance of 'SDR' and 'SDR+GR' are almost the same at different levels of energy harvesting.

\section{Conclusion}
In this paper, we have considered the design of the transmit and AN covariance matrices for a MISO multicasting SWIPT system with multiple-antenna eavesdroppers. In particular, the secrecy rate maximization problem was solved with transmit power and energy harvesting constraints. The original problem was not convex in terms of the covariance matrices. To overcome this non-convexity issue, we convert the original problem into a convex one by exploiting semidefinite relaxation and \emph{Charnes-Cooper} transformation. In addition, we used GR to construct the rank one solution for the original problem from the non-rank one results. Simulation results were provided to validate the performance of proposed schemes.
\begin{appendix}
	\subsection*{Proof of Proposition \ref{proposition:rank_proof}}\label{proof_of_proposition}
Let the optimal value obtained by solving (21) to be $\tilde{f}^{*}(t)$. We consider the following minimization problem. 
\begin{align}
&\min_{\mathbf{Q}_{s},\mathbf{V},\rho_{s,k}} ~\textrm{tr}(\mathbf{Q}_{s}) \nonumber\\
&s.t. ~ \textrm{tr}(\mathbf{Q}_{s})+\textrm{tr}(\mathbf{V})\leq P_{total}\nonumber\\
&\mathbf{h}^{H}_{s,k}[\mathbf{Q}_{s}-\tilde{f}^{*}(t)\mathbf{V}]\mathbf{h}_{s,k}-\tilde{f}^{*}(t)(\sigma^{2}_{sa,k}+\frac{\sigma^{2}_{sp,k}}{\rho_{s,k}})\geq 0,\forall k,\nonumber\\
&(t^{-1}-1)(\mathbf{H}^{H}_{e,l}\mathbf{V}\mathbf{H}_{e,l}+\sigma^{2}_{e,l}\mathbf{I})-\mathbf{H}^{H}_{e,l}\mathbf{Q}_{s}\mathbf{H}_{e,l}\succeq 0,\forall l,\nonumber\\
&\mathbf{h}^{H}_{s,k}(\mathbf{Q}_{s}+\mathbf{V})\mathbf{h}_{s,k}\geq \frac{\bar{E}_{s}}{1-\rho_{s,k}}-\sigma^{2}_{sa,k},\forall k,\nonumber\\
&\textrm{tr}[\mathbf{H}^{H}_{e,l}(\mathbf{Q}_{s}+\mathbf{V})\mathbf{H}_{e,l}]\geq \bar{E}_{e}-N_{E}\sigma^{2}_{ea,l},\forall l,\nonumber\\
&0<\rho_{s,k}\leq 1, \mathbf{Q}_{s}\succeq 0, \mathbf{V}\succeq 0.  
\end{align}
The optimal solution of the problem in (21) is also the optimal solution of the problem in (23) and vice versa \cite{li2011}. Therefore, we can analyse the rank property of the optimal solution of the problem in (21) by analysing that of the problem in (23). First, we write the Lagrange dual function as
\begin{align}
&\mathcal{L}(\mathbf{Q}_{s},\mathbf{V},\mathbf{Z},\mathbf{Y},\lambda,\mu_{k},\mathbf{A}_{l},\alpha_{k},\beta_{l})\!=\!\textrm{tr}(\mathbf{Q}_{s})\!-\!\textrm{tr}(\mathbf{Z}\mathbf{Q}_{s})\!-\!\textrm{tr}(\mathbf{Y}\mathbf{V})\nonumber\\
&\!-\!\lambda[\textrm{tr}(\mathbf{Q}_{s})\!+\!\textrm{tr}(\mathbf{V})\!-\!P_{total}]\!-\!\sum_{k=1}^{K}\mu_{k}\bigg[\mathbf{h}^{H}_{s,k}[\mathbf{Q}_{s}\!-\!\tilde{f}^{*}(t)\mathbf{V}]\mathbf{h}_{s,k}\nonumber\\
&\!-\!\tilde{f}^{*}(t)(\sigma^{2}_{sa,k}\!+\!\frac{\sigma^{2}_{sp,k}}{\rho_{s,k}})\bigg]\!-\!\sum_{l=1}^{L}\textrm{tr}\{\mathbf{A}_{l}[(t^{-1}\!-\!1)(\mathbf{H}^{H}_{e,l}\mathbf{V}\mathbf{H}_{e,l}\!+\!\sigma^{2}\mathbf{I})\nonumber\\
&\!-\!\mathbf{H}^{H}_{e,l}\mathbf{Q}_{s}\mathbf{H}_{e,l}]\}\!-\!\sum_{k=1}^{K}\alpha_{k}[\mathbf{h}^{H}_{s,k}(\mathbf{Q}_{s}\!+\!\mathbf{V})\mathbf{h}_{s,k}\!-\!\frac{\bar{E}_{s}}{1-\rho_{s,k}}\!+\!\sigma^{2}_{sa,k}]\nonumber\\
&-\sum_{l=1}^{L}\beta_{l}\bigg[\textrm{tr}[\mathbf{H}^{H}_{e,l}(\mathbf{Q}_{s}+\mathbf{V})\mathbf{H}_{e,l}]-\bar{E}_{e}+N_{E}\sigma^{2}_{ea,l}\bigg]
\end{align}
where $\mathbf{Z}\in\mathbb{H}_{+}^{N_{T}}$, $\mathbf{Y}\in\mathbb{H}_{+}^{N_{T}}$, $\lambda\in\mathbb{R}_{+}$,$\mu_{k}\in\mathbb{R}_{+}$ $\mathbf{A}_{l}\in\mathbb{H}_{+}^{N_{E}}$, $\alpha_{k}\in\mathbb{R}_{+}$ and $\beta_{l}\in\mathbb{R}_{+}$ are Langrangian dual vaiables associated with (23). Then we derive the following \emph{Karush-Kuhn-Tucker} (KKT) conditions \cite{boyd2004convex}:
\begin{subequations}\label{eq:KKT_conditions}
	\begin{align}
	&\frac{\partial\mathcal{L}}{\partial\mathbf{Q}_{s}}=\mathbf{I}-\mathbf{Z}+\lambda\mathbf{I}-\sum_{k=1}^{K}\mu_{k}\mathbf{h}_{s,k}\mathbf{h}^{H}_{s,k}+\sum_{l=1}^{L}\mathbf{H}_{e,l}\mathbf{A}_{l}\mathbf{H}^{H}_{e,l}\nonumber\label{eq:KKT1}\\
	&-\sum_{k=1}^{K}\alpha_{k}\mathbf{h}_{s,k}\mathbf{h}^{H}_{s,k}-\sum_{l=1}^{L}\beta_{l}\mathbf{H}_{e,l}\mathbf{H}^{H}_{e,l}=0,\\
	&\frac{\partial\mathcal{L}}{\partial\mathbf{V}}\!=\!\!-\!\mathbf{Y}\!+\!\lambda\mathbf{I}\!-\!\sum_{k=1}^{K}\mu_{k}\tilde{f}^{*}(t)\mathbf{h}_{s,k}\mathbf{h}^{H}_{s,k}\!-\!\sum_{l=1}^{L}(\frac{1}{t}\!-\!1)\mathbf{H}_{e,l}\mathbf{A}_{l}\mathbf{H}^{H}_{e,l}\nonumber\label{eq:KKT2}\\
	&-\sum_{k=1}^{K}\alpha_{k}\mathbf{h}_{s,k}\mathbf{h}^{H}_{s,k}-\sum_{l=1}^{L}\beta_{l}\mathbf{H}_{e,l}\mathbf{H}^{H}_{e,l}=0,\\
	&\mathbf{Z}\mathbf{Q}_{s}=0, \mathbf{Y}\mathbf{V}=0, \mathbf{Z}\succeq 0, \mathbf{Y}\succeq 0.\label{eq:KKT3}
	\end{align}
\end{subequations}
The following equality holds:
\begin{align}
&\eqref{eq:KKT1}-\eqref{eq:KKT2}=\mathbf{I}-\mathbf{Z}+\mathbf{Y}-\sum_{k=1}^{K}\mu_{k}[1+\tilde{f}^{*}(t)]\mathbf{h}_{s,k}\mathbf{h}^{H}_{s,k}+\nonumber\\
&\sum_{l=1}^{L}t^{-1}\mathbf{H}_{e,l}\mathbf{A}_{l}\mathbf{H}^{H}_{e,l}=0,\nonumber\\
&\rightarrow\! \mathbf{Z}\!=\!\mathbf{I}\!+\!\mathbf{Y}\!+\!\sum_{l=1}^{L}t^{-1}\mathbf{H}_{e,l}\mathbf{A}_{l}\mathbf{H}^{H}_{e,l}\!-\!\sum_{k=1}^{K}\mu_{k}[1\!+\!\tilde{f}^{*}(t)]\mathbf{h}_{s,k}\mathbf{h}^{H}_{s,k}\nonumber\\
&\rightarrow[\mathbf{I}+\mathbf{Y}+\sum_{l=1}^{L}t^{-1}\mathbf{H}_{e,l}\mathbf{A}_{l}\mathbf{H}^{H}_{e,l}]\mathbf{Q}_{s}\nonumber\\
&=[1+\tilde{f}^{*}(t)](\sum_{k=1}^{K}\mu_{k}\mathbf{h}_{s,k}\mathbf{h}^{H}_{sk})\mathbf{Q}_{s}
\end{align}
From the above equality, the following rank relation can be derived:
\begin{align}
&\textrm{rank}(\mathbf{Q}_{s})=\textrm{rank}\{[1+\tilde{f}^{*}(t)][\mathbf{I}+\mathbf{Y}+\sum_{l=1}^{L}t^{-1}\mathbf{H}_{e,l}\mathbf{A}_{l}\mathbf{H}^{H}_{e,l}]^{-1}\nonumber\\
&(\sum_{k=1}^{K}\mu_{k}\mathbf{h}_{s,k}\mathbf{h}^{H}_{s,k})\mathbf{Q}_{s}\}\leq\textrm{rank}(\sum_{k=1}^{K}\mu_{k}\mathbf{h}_{s,k}\mathbf{h}^{H}_{s,k})\leq K.
\end{align}
In order to derive this rank condition, the following $lemma$ is required \cite{huang2010rank}.\\
$Lemma$ $1$: Consider the following SDP problem
\begin{align}
&\min_{\mathbf{W}_{k}\in\mathbb{H}^{N},k=1,...,K} ~\sum_{k=1}^{K}\textrm{tr}(\mathbf{A}_{k}\mathbf{W}_{k})\nonumber\\
&s.t.~\sum_{k=1}^{K}\textrm{tr}(\mathbf{B}_{m,k}\mathbf{W}_{k})\unrhd_{m} b_{m},m\!=\!1,...,M,\mathbf{W}_{k}\succeq0, k\!=\!1,...,K,
\end{align}
where $b_{m}\in\mathbb{R}$, $\mathbf{A}_{k}$, $\mathbf{B}_{m,k}\in\mathbb{H}^{N}$, and for each $m$, $\unrhd_{m} \in \{\geq,=,\leq\}$. Provided that the problem in (28) is feasible, then there exists an optimal solution $(\mathbf{W}^{*}_{1},...,\mathbf{W}^{*}_{K})$, such that $\sum_{k=1}^{K}\textrm{rank}^{2}(\mathbf{W}^{*}_{k})\leq M$.
\\By applying $Lemma$ 1 to be the problem in (23), we have the conclusion that exists an optimal $\mathbf{Q}_{s}$ and an optimal $\mathbf{V}$ that $\textrm{rank}^{2}(\mathbf{Q}_{s})+\textrm{rank}^{2}(\mathbf{V})\leq 2K+L$.	

\end{appendix}



\bibliographystyle{IEEEtran}
\bibliography{referenceIEEE}
%

\end{document}